\documentclass{article}

\usepackage{arxiv}

\usepackage{mathptmx}
\usepackage{soul}\setuldepth{article}
\usepackage{enumitem}
\usepackage{times}
\normalfont
\usepackage[T1]{fontenc}
\usepackage{booktabs}
\usepackage{fancyvrb}
\usepackage[table,xcdraw]{xcolor}
\usepackage{makecell}
\usepackage{booktabs}
\usepackage{footmisc}
\usepackage{graphicx}
\usepackage{listings}
\usepackage{algorithm}
\usepackage{algorithmic}
\usepackage{changepage}
\usepackage{url}
\usepackage{color}
\usepackage[htt]{hyphenat}
\usepackage[square,numbers,sort]{natbib} % Required for adding bibliography

\usepackage[
bookmarks=true,
hypertexnames=true,
hyperindex=true,
hyperfootnotes=true,
linktocpage=false,
hyperfigures=true,
breaklinks=true, 
pdfborder={0 0 0},
colorlinks=true,
linkcolor=black,
unicode=true,						
citecolor=blue,
urlcolor=blue,
pdfinfo={
Title={RDFGraphGen: An RDF Graph Generator based on SHACL Shapes},
Author={Milos Jovanovik, Marija Vecovska, Maxime Jakubowski, Katja Hose},
Keywords={Data Generator, Synthetic Data, Knowledge Graphs, RDF, SHACL}
}
]{hyperref}
\usepackage{fancyvrb}
\usepackage[title]{appendix}

\title{RDFGraphGen: An RDF Graph Generator based on SHACL Shapes}

\author{
  \textbf{Milos Jovanovik\textsuperscript{1,2}, Marija Vecovska\textsuperscript{2}, Maxime Jakubowski\textsuperscript{1}, Katja Hose\textsuperscript{1}}\\
  \textsuperscript{1} Institute of Logic and Computation, TU Wien, Austria\\
  \textsuperscript{2} Faculty of Computer Science and Engineering\\
  Ss. Cyril and Methodius University in Skopje, N. Macedonia\\
  \texttt{\{name.surname\}@tuwien.ac.at}\\
  \texttt{\{name.surname\}@finki.ukim.mk}\\
}

\definecolor{codegray}{rgb}{0.5,0.5,0.5}

\newtheorem{example}{Example}

\def\hb{\hbox to 11.5 cm{}}

\begin{document}

\maketitle

\begin{abstract}

Developing and testing modern RDF-based applications often requires access to RDF datasets with certain characteristics. Unfortunately, it is very difficult to publicly find domain-specific knowledge graphs that conform to a particular set of characteristics. Hence, in this paper we propose RDFGraphGen, an open-source RDF graph generator that uses characteristics provided in the form of SHACL (Shapes Constraint Language) shapes to generate synthetic RDF graphs. RDFGraphGen is domain-agnostic, with configurable graph structure, value constraints, and distributions. It also comes with a number of predefined values for popular schema.org classes and properties, for more realistic graphs. Our results show that RDFGraphGen is scalable and can generate small, medium, and large RDF graphs in any domain.
\end{abstract}

%%%%%%%%%%%%%%%%%%%%%%%%%%%%%%%%%%%%%%%%%%

\section{Introduction}
\label{sec:introduction}

The acceptance of RDF~\cite{RDF} as a standard for knowledge representation, publication on the Web~\cite{webdatacommons2024}, and internally in organizations~\cite{buchgeher2021knowledge}, has led to the development of many tools, libraries, and applications that work with or use RDF as a data model. During the development lifecycle of each of these, benchmarking and testing are important steps, which validate the usage scenarios and provide metrics about their real-world usability~\cite{angles2014linked}. These benchmarks and test scenarios often require application-specific or domain-specific RDF datasets with certain characteristics, that are not always readily available. One possible solution to this is synthetic RDF datasets --- datasets containing entities and data that are artificial but follow the structure, distribution, and vocabulary expected from applications or systems. They fulfill the need for data in a specific format, for a specific domain, application, or system to be tested or benchmarked~\cite{10.1007/978-3-030-00072-1_1,10.1007/978-3-319-69146-6_1}. Depending on the specific needs, these synthetic RDF datasets can be small or consist of millions of different entities~\cite{duan2011apples}.

One of the more recent additions to the technology stack around RDF is the Shapes Constraint Language (SHACL)~\cite{SHACL}. Its main purpose is defining structural constraints on RDF graphs, called \emph{shapes}. SHACL is designed to validate RDF graphs to ensure their data quality. However, these shapes can also be used as \emph{blueprints} for generating domain-agnostic RDF data. Recognizing this potential and aiming to address the need for an RDF data generator, this paper presents RDFGraphGen. RDFGraphgen is, to the best of our knowledge, the first \emph{SHACL-based} RDF data generation tool. As opposed to the usual task of \emph{validating} RDF data, we use SHACL shapes for \emph{generating} RDF data.

Hence, in this paper we describe the design and implementation of RDFGraphGen --- a domain-independent generator of RDF graphs based on SHACL shapes. SHACL shapes are interpreted as a description of the desired structure of an RDF graph, and in accordance with this view the corresponding RDF triples are generated. RDFGraphGen is domain-agnostic: the source SHACL shapes can be from any domain. The generator can generate an RDF graph with a specified scale factor that determines the number of generated entities, providing flexibility for end-users. RDF data generated by \mbox{RDFGraphGen} can be used for benchmarking, testing, quality control, or other similar tasks in the application lifecycle of systems using RDF. Additionally, RDFGraphGen is open-source\footnote{RDFGraphGen on GitHub: \url{https://github.com/etnc/RDFGraphGen/}\label{fn:rdfgraphgen-github}} and available as a ready-to-use Python package\footnote{RDFGraphGen on PyPi: \url{https://pypi.org/project/rdf-graph-gen/}\label{fn:rdfgraphgen-pypi}}, under the MIT license.

This paper is organized as follows. Section~\ref{sec:relatedwork} reviews related work and existing tools. Section~\ref{sec:rdfgraphgen-overview} describes the design and the underlying algorithms of RDFGraphGen. Section~\ref{sec:examples} illustrates its features, usability, and data generation through a practical example. Section~\ref{sec:performance-evaluation} outlines the results of the performance evaluation of RDFGraphGen. Finally, Section~\ref{sec:conclusion} concludes with summary and future work.

%%%%%%%%%%%%%%%%%%%%%%%%%%%%%%%%%%%%%%%%%%

\section{Related Work}
\label{sec:relatedwork}

Generating synthetic RDF data is not a new topic as it has been of interest to the research community for quite a long time. In this section, we provide an overview of existing generators of synthetic RDF datasets. 

Tab2KG \cite{gottschalk2022tab2kg} is a method that is used for interpretation of tables with previously unseen data and automatically infers their semantics to transform them into semantic data graphs. The Tab2KG algorithm transforms tabular data into a semantic data graph by automatically inferring the domain ontology and mapping the table columns to the ontology classes and properties, before transforming the rows of the table into RDF triples. In short, the tabular data is provided, and the ontology is adjusted to fit this data. In contrast, RDFGraphGen uses explicit rules and constraints --- it takes a description of a target data graph as a SHACL shapes graph and generates entities according to this description, using random or structured values for the objects in the RDF triples of the generated entities.

GAIA \cite{raynaud2016generic} is a generic RDF data generator that allows users to generate RDF triples by conforming to an ontology. It is OWL-based and generates RDF objects based on any correctly defined OWL ontology. The generator correctly generates a user-defined number of objects following the OWL ontology but offers no way to constrain the objects' values beyond datatype. RDFGraphGen uses a SHACL shapes graph as a description of the entities that should be generated, allowing the user to describe the object's values in great detail. RDFGraphGen also allows using properties from multiple ontologies in a generated entity since it does not generate data based on a specific ontology. The ontology is implicitly defined in SHACL shapes via the entity classes and the specified properties.

GRR \cite{blum2010generating} is a system for generating random RDF data using SPARQL-like syntax to describe the desired ontology. GRR can generate entities using the desired ontology and allows the user to provide input for the objects' values in the triples. However, GRR offers no method for constraining these values beyond providing them beforehead, making the process much more complicated for the user. RDFGraphGen in contrast uses the input SHACL shapes graph to constrain the values of the objects in the generated RDF triples.

PyGraft \cite{hubert2024pygraftconfigurablegenerationsynthetic} is another tool designed for the creation of highly customized, domain-independent RDF datasets and RDF schemas. It takes input RDF schemas and uses a set of rules and constraints to generate synthetic RDF triples that are compliant with the input structure. It has the ability to simulate realistic data distributions, including edge and node properties, while ensuring that the generated data adheres to the input ontology's constraints. PyGraft is scalable, which makes it suitable for large datasets, and it is capable of handling various RDF vocabularies to produce diverse datasets that closely mimic real-world RDF data. RDFGraphGen, on the other hand, relies on the level of details and constraints contained in the input SHACL shapes graph to generate the desired synthetic RDF knowledge graph.

GenACT \cite{singh2024genact} is a data generator for temporal and evolving RDF graphs in the domain of social media activity during academic events --- more specifically, academic conference tweets (ACT). This approach tries to address the gap of missing real-world knowledge graphs with specific capabilities that are necessary for the use cases in this domain. In contrast, RDFGraphGen is domain-independent and tries to solve the same gap from a more generalized perspective.

%%%%%%%%%%%%%%%%%%%%%%%%%%%%%%%%%%%%%%%%%%

\section{RDFGraphGen}
\label{sec:rdfgraphgen-overview}

In this section, we describe the design of RDFGraphGen, we explain how the graph structure and the literal values are generated and how to use the tool to generate RDF data from SHACL shapes.

\begin{figure}[!ht]
    \centering
    \includegraphics[width=0.75\linewidth]{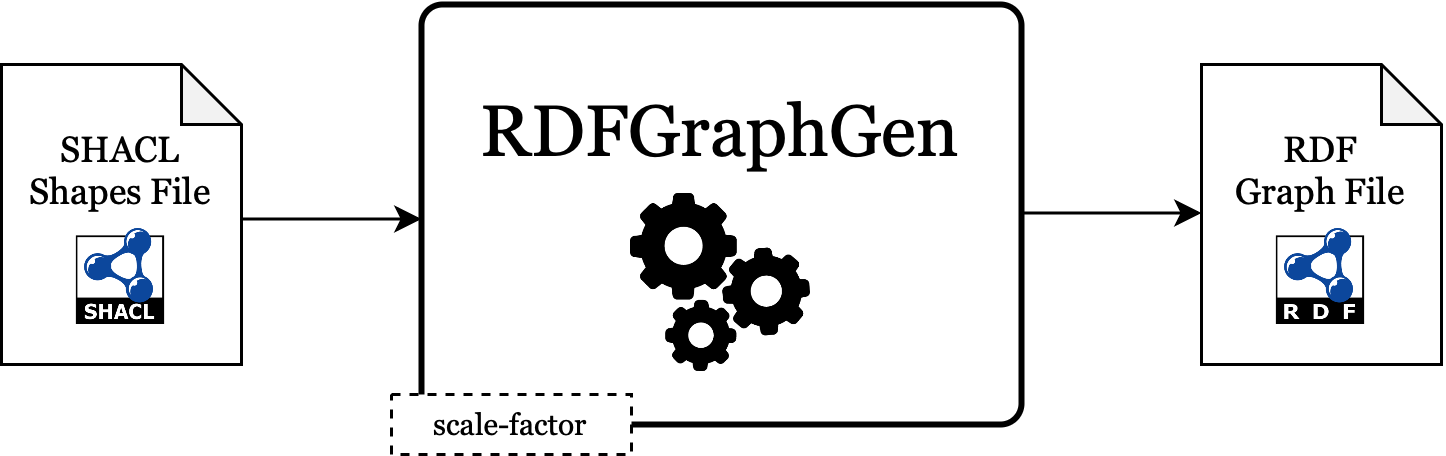}
    \caption{RDFGraphGen Workflow Diagram}
    \label{fig:rdfgraphgen-diagram}
\end{figure}

RDFGraphGen has a straightforward workflow (Figure \ref{fig:rdfgraphgen-diagram}). Given a SHACL shapes graph as input, RDFGraphGen outputs RDF data. The size of the generated RDF graph is controlled by an input parameter: \emph{scale-factor}. We also explain how the scale factor affects the number of entities and RDF triples. 

\subsection{Graph Generation Algorithm}
\label{sec:algorithm}

The goal is to generate realistic and useful RDF data. For the input shapes graphs, \mbox{RDFGraphGen} generates a data graph that conforms to them. The aim is to create an algorithm that uses the structure of the input SHACL shapes and generates data following this structure. The goal is to use as many details from the shapes graph in order to populate the generated RDF data graph, even if the shapes are contradictory. Generally, the task of generating conforming data graphs is non-trivial\footnote{Because SHACL corresponds to a logic~\cite{shacldl}, generating data corresponds to finding an model for a given SHACL theory. Even for perceived ``simple'' shapes, the task might be best solved by resorting to first-order logical reasoners.}.

\subsubsection{Preliminaries on SHACL} 

To provide sufficient background, we start by introducing SHACL \cite{SHACL}. In general, the main purpose of SHACL is to define a set of constraints on nodes called \emph{shapes}, which in turn are used to define constraints on an RDF graph. The set of shapes is referred to as a \emph{shapes graph}\footnote{It is called a \emph{graph} because SHACL syntax is defined in RDF itself.} while the data that is checked is called the \emph{data graph}. When the data graph satisfies all the shapes from the shapes graph, we say it \emph{conforms}. A node in the data graph that must satisfy a shape is called the \emph{focus node}. Focus nodes for a given shape are specified by \emph{target declarations}.

Conceptually, there are two different types of shapes: \emph{node shapes}, which describe constraints on the focus nodes themselves, and \emph{property shapes} describing constraints on a set of values reachable from the focus node by a specified property. Structurally, a node shape $s=(n,C,T)$ consists of three components: (1) a shape name $n$, typically an IRI, (2) a conjunction of constraint components $C$, and (3) a (possibly empty) set of target declarations $T$. When there are target declarations, the shape can be considered a constraint on the data graph. Similarly, a property shape $p=(n,C,T,E)$ consists of all the components of a node shape, with an additional element: the \emph{property path} $E$. For our purposes, the property path is always a predicate path, i.e., an IRI.

A \emph{constraint component} describes the specific conditions that a focus node must satisfy. Node shapes can be considered as a conjunction of constraint components --- all constraint components must be satisfied for the focus node. In particular, shape-based constraint components\footnote{\url{https://www.w3.org/TR/shacl/\#core-components-shape}} allow for creating a network of shapes that can be used to inform the structure of the generated data graph.

\newcommand{\generatefromnodeshape}{\textsc{GenerateFromNodeshape}}
\newcommand{\generatefrompropshape}{\textsc{GenerateFromPropshape}}
\newcommand{\determinecardinality}{\mathit{determineCardinality}}
\newcommand{\createschemavalues}{\mathit{createSchemaValues}}
\newcommand{\createliteralvalues}{\mathit{createLiteralValues}}
\newcommand{\generatelogical}{\textsc{GenerateLogical}}

\subsubsection{Generating the Graph}

The algorithm for generating RDF data is primarily based on the set of constraint components. The shape name and the target declarations are only used as additional information. We assume that the shapes graph is well-formed\footnote{\url{https://www.w3.org/TR/shacl/\#dfn-well-formed}} and non-recursive\footnote{\url{https://www.w3.org/TR/shacl/\#shapes-recursion}}, i.e., that it is syntactically correct SHACL and there are no self-referential shapes.

The algorithm is designed as follows. Given an input SHACL shapes graph $S$, and the number of desired entities $m$, RDFGraphGen generates an RDF graph $G$ that contains $m$ entities per explicitly defined node shape $s$ in $S$. Specifically, we generate $m$ entities for each node shape $s$, such that the triple $(s,$ $\mathsf{rdf{:}type},$ $\mathsf{sh{:}NodeShape})$ appears in the shapes graph $S$.

\begin{algorithm}[!ht]
\caption{$\generatefromnodeshape(S, s, e)$}
\begin{algorithmic}[1]
    \STATE $G = \emptyset$
    \STATE $s = (s_n, C, s_T)$
    \FOR{each constraint component $c\in C$}
    \IF{$c$ is a property shape component}
    \STATE Let $p$ be the property shape associated with $c$.
    \STATE $G', E := \generatefrompropshape(S,p)$
    \STATE Let $r$ the predicate path associated with $p$.
    \STATE $G := G \cup G' \cup \{(e,r,e') \mid e'\in E\}$
    \ELSIF{$c$ is a node constraint component}
    \STATE Let $s'$ be the shape from $S$ associated with $c$.
    \STATE $G := G \cup \generatefromnodeshape(S,s',e)$
    \ELSIF{$c$ is a logical constraint component}
    \STATE $G := G\cup \generatelogical(S,c,e)$
    \ENDIF
    \ENDFOR
    \RETURN $G$
\end{algorithmic}
\label{alg:generatefromnodeshape_maxime}
\end{algorithm}

\begin{algorithm}[!ht]
    \caption{$\generatefrompropshape(S,p)$}
    \begin{algorithmic}[1]
    \STATE $G := \emptyset$
    \STATE $p = (p_n, C, p_T, q)$
    \STATE $\mathsf{card} := \determinecardinality(C)$
    \IF{$q$ is a supported schema.org predicate}
    \RETURN $G, 
    \createschemavalues(q, \mathsf{card})$
    \ELSIF{$\exists c \in C$ such that $c$ is a datatype constraint component}
    \STATE Let $\mathsf{datatype}$ be the datatype associated with $c$.
    \RETURN $G, \createliteralvalues(\mathsf{datatype},\mathsf{card})$
    \ENDIF
    \STATE Let $E := \{e_1,\dots,e_\mathsf{card}\}$ be a set of newly created IRIs.
    \IF{$\exists c\in C$ such that $c$ is a property shape component}
    \STATE Let $C'\subseteq C$ be the set of property shape components from $C$.
    \STATE Let $S'$ be the set of all property shapes associated with $C'$.
    %\STATE Let $S'=\{s'_1,\dots s'_k\}$ be the set of all $k$ property shapes associated with $C'$.
    %\STATE Partition $E$ into disjoint sets $E_1,\dots,E_k$
    %\FOR{each $s_i'\in S'$}
    \FOR{each $s'\in S'$}
    \STATE Let $q$ be the predicate path associated with $s'$.
    \STATE $G', E' := \generatefrompropshape(S,s')$
    \STATE $G := G \cup G' \cup \bigcup_{e\in E}\{(e,q,e')\mid e'\in E'\}$
    \ENDFOR
    %\STATE $G := \bigcup_{e\in E}\bigcup_{s\in S'}\textsc{GenerateFromNodeshape}(S,s,e)$
    \ELSIF{$\exists c\in C$ such that $c$ is a node constraint component}
    \STATE Let $C'\subseteq C$ be the set of node constraint components from $C$.
    % \STATE Let $S'$ be the set of node shapes associated with $C'$.
    \STATE Let $S'=\{s'_1,\dots,s'_k\}$ be the set of all $k$ node shapes associated with $C'$.
    \STATE Partition $E$ into mutually disjoint sets $E_1,\dots,E_k$
    \FOR{each $s'_i\in S'$}
    \STATE $G= G \cup\bigcup_{e\in E_i}\generatefromnodeshape(S,s'_i,e)$
    \ENDFOR
    \ELSIF{$\exists c\in C$ such that $c$ is a logical constraint component}
    \STATE Let $C\subseteq C'$ be the set of logical constraint components from $C$.
    \STATE $G := \bigcup_{c'\in C'}\bigcup_{e\in E}\generatelogical(S,c',e)$
    \ENDIF
    \RETURN $G,E$
    \end{algorithmic}
    \label{alg:generatefrompropshape_maxime}
\end{algorithm}

\begin{algorithm}[!ht]
    \caption{$\generatelogical(S,c,e)$}
    \begin{algorithmic}[1]
    \STATE $S' := \emptyset$
    \IF{$c$ is an `and' constraint component}
    \STATE $S'$ is the set of shapes from $S$ associated with $c$.
    \ELSIF{$c$ is a `or' constraint component}
    \STATE $S'$ is a random, non-empty subset of the shapes in $S$ associated with $c$.
    \ELSIF{$c$ is a 'xone' constraint component}
    \STATE $S'$ is a set containing exactly one random shape from $S$ associated with $c$.
    \ENDIF
    \RETURN $\bigcup_{s'\in S'} \generatefromnodeshape(S,s',e)$
    \end{algorithmic}
    \label{alg:generatelogical_maxime}
  \end{algorithm}

We focus our explanation on the generation of a single entity $e$ for a node shape $s$. Recall that a shape contains a set of constraint components $C$. The procedure $\generatefromnodeshape(S,s,e)$ is described in Algorithm~\ref{alg:generatefromnodeshape_maxime}. It uses the shapes graph $S$, a node shape $s$, and an IRI $e$ to generate triples describing entity $e$ using $s$, possibly referring to other shapes from $S$, and outputs a synthetic RDF graph $G$. It looks at every supported constraint component and generates data accordingly. For node constraint components (\texttt{sh:node}), it simply generates some data for every component, for entity $e$. Similarly, for the supported logical constraint components \texttt{sh:and}, \texttt{sh:or}, and \texttt{sh:xone}, we select an approprate subset of shapes for which data is generated w.r.t. $e$. (see Algorithm~\ref{alg:generatelogical_maxime}).

\begin{example} \label{ex:myshapea}
  Consider the shape \texttt{<MyShape>}:
\begin{verbatim}
<myShapeA> a sh:NodeShape ;
  sh:node <myShapeB>, <myShapeC> ;
  sh:or ( <myShapeD> <myShapeE> ) .
\end{verbatim}
  %  sh:property [ sh:path <pred> ; sh:minCount 1 ] ;
  where we assume \texttt{<myShapeB>}, \dots,\texttt{<myShapeE>} are
  defined. The generator creates the IRI $e$ and calls
  $\generatefromnodeshape(S,s,e)$ for $s=\texttt{<myShapeA>}$
  (initially), $s=\texttt{<myShapeB>}$ and $s=\texttt{<myShapeC>}$
  (all node constraint components), and also, randomly, for
  $s=\texttt{<myShapeD>}$ or $s=\texttt{<myShapeE>}$ or both
  (logical constraints).
\end{example}

Finally, and most importantly, data is generated for every associated property shape (\texttt{sh:property}) $p$, handled by the procedure $\generatefrompropshape(S,p)$ described in Algorithm~\ref{alg:generatefrompropshape_maxime}, which returns a set of new RDF terms $E$, and (a possibly empty) RDF graph $G'$ representing triples that are recursively generated for entities from $E$. This is the most crucial procedure, as the property shapes indicate the structure of the generated RDF graph by describing the immediate neighborhood.

\begin{example} \label{ex:myshapea2}
  Continuing from Example~\ref{ex:myshapea}, we can add the following
  additions to the shape definition:
\begin{verbatim}
<myShapeA> sh:property <myPropShape> .
<myPropShape> sh:path <pred> ; sh:minCount 1 .
\end{verbatim}
  This means that the generator calls $\generatefrompropshape(S,p)$,
  where
  $p=(\texttt{<myPropShape>}, \{\texttt{sh:minCount 1}\}, \emptyset,
  \texttt{<pred>})$
\end{example}

The procedure works as follows. First, we determine how many values are generated for the property shape (indicated by $\determinecardinality(C)$) based on the keywords \texttt{sh:minCount} and \texttt{sh:maxCount}. Essentially, a random integer is selected between the minimum and maximum cardinality. If there is an inconsistency (i.e., the minimal cardinality is greater than the maximal), the minimal indicated cardinality is followed.

Next, we distinguish between generating literal values, or new entities. If the predicate path is a schema.org predicate, we generate an appropriate value with the helper function $\createschemavalues(q,\mathtt{card})$. Essentially, $\texttt{card}$ indicates the number of values to be generated appropriate to $q$. Similarly, if the values are intended to be of a certain datatype (by default \texttt{xsd:string} values), these are handled accordingly by the helper function $\createliteralvalues(\mathtt{datatype},\mathtt{card})$. We elaborate on both helper functions in Section~\ref{sec:literalvalues}.

Finally, if we expect the values for the property to be other entities, e.g., referring to other shapes with \texttt{sh:node}, \texttt{sh:property}, or a logical constraint component, then the generation procedure is called recursively in accordance with the indicated structure of the shape.

\begin{example}
  Recall \texttt{<myPropShape>} from Example~\ref{ex:myshapea2}. The
  generator recognizes that at least one value must be generated, so
  it will choose $\texttt{card} \geq 1$, for example
  $\texttt{card} = 2$. Next, because \texttt{<pred>} is not a
  specially supported predicate path and because there is no datatype
  explicitly defined, it will generate two random strings
  $\texttt{str}_1$ and $\texttt{str}_2$, and adds the triples ($e$,
  \texttt{<pred>}, $\texttt{str}_1$) and ($e$, \texttt{<pred>},
  $\texttt{str}_2$) to the output graph.
\end{example}

\subsubsection{Size of the Generated RDF Graph}

RDFGraphGen uses a scale factor to determine the size of the generated RDF dataset. The value of the parameter corresponds to the number of entities that will be generated from all top-level node shapes in the input SHACL graph. By \emph{top-level} node shapes we mean node shapes that are either stand-alone and not related to other node shapes from the SHACL graph, or that only appear as subjects in the predicates pointing to other node shapes from the SHACL graph. The other node shapes, which solely appear as objects in predicates from other node shapes, are generated in numbers depending on the cardinality of the same predicate, as described in Algorithm~\ref{alg:generatefrompropshape_maxime}.

\begin{example} \label{ex:myshapea4}
  Consider the shape \texttt{<myShapeF>}:
\begin{verbatim}
<myShapeF> a sh:NodeShape ;
    sh:property [ sh:path <predicate> ;
        sh:node <myShapeG> ;
        sh:minCount 2 ;
        sh:maxCount 4 ] .
\end{verbatim}
  This means that \texttt{<myShapeF>} is a top-level shape, connected via \texttt{<predicate>} to the non top-level shape \texttt{<myShapeG>}. 
  
  If the scale-factor is set to 100, the generator will generate 100 entities based on \texttt{<myShapeF>}, and between 200--400 entities based on \texttt{<myShapeG>} (i.e., 2 to 4 per entity based on \texttt{<myShapeF>}), resulting in a total of 300--500 entities in the generated RDF graph.
\end{example}

\subsection{Generating Literal Values}
\label{sec:literalvalues}

To generate \emph{useful} RDF data, we generate meaningful literal values whenever possible. Recall that Algorithm~\ref{alg:generatefrompropshape_maxime} explicitly refers to schema.org for the creation of values, based on the predicate path of the property shape. The focus on schema.org is not a fundamental limitation: our implementation can easily be extended to include other vocabularies in the future.

If we need to generate a value that is not related to schema.org, then the values are are generated based on the specified datatype (\texttt{sh:datatype}), and if none is given, we default to \texttt{xsd:string}.

\subsubsection{Generating Schema.org Literals} 

When generating a new literal, the first step is to try and infer what would be its logical value, given the predicate and the RDF type of the entity as defined in the input SHACL shape. For example, if a triple is being generated for an entity of type \texttt{schema:Person} from the widely used schema.org vocabulary \cite{guha2016schema}, where the predicate is \texttt{schema:firstName}, it is natural that the generated value of the object should be a human first name. In this case, the generator uses a set of first names to pick one randomly. Even further, if the \texttt{schema:Person} entity has a \texttt{schema:gender} value of either male or female, the generator will specifically use the list of male or female first names, accordingly, to pick one at random. 

The generator contains sets of collected values for the most common properties from schema.org. These sets are used by RDFGraphGen to randomly select a value for such properties, when they appear in the SHACL shapes. More specifically, RDFGraphGen has collections of male and female names and surnames, job titles, addresses and street names, book titles, book genres, TV series titles, movie titles, movie genres, etc., that have been collected from online resources. This feature of RDFGraphGen can easily be extended in the future to include many more domains.

Additionally, the generator contains several rules about constructing values for other well-known schema.org predicates, e.g., a full name or an email address. In scenarios where the input SHACL shape uses \texttt{schema:name}, the value is constructed by concatenating a random (gendered) first name and a last name. When the predicate \texttt{schema:email} is used without a specified pattern, RDFGraphGen generates an email address in the form \texttt{firstname-lastname@gmail.com}.

If the predicate is unknown, it can still provide useful information to determine a suitable value. For example, if the name of the predicate is \texttt{birthDate}, from a random or unknown vocabulary, the term \texttt{date} suggests that the object should be a date. This will be caught by RDFGraphGen and a date value will be generated. Similar behavior is implemented for other predicates from unknown vocabularies, such as \texttt{telephone}, \texttt{phone}, \texttt{email}, etc.

\subsubsection{Generating Generic Values} 

When the value cannot be picked from a pre-defined set of specific values, we generate it randomly. This random generation step uses all of the constraints components defined in the input SHACL shape, e.g., datatype, minimum length, maximum length, pattern, minimum and maximum value, etc.

This is what makes the RDFGraphGen generator general-purpose and domain-independent: it can generate synthetic RDF triples and graphs from any domain (life sciences, social networking, entertainment, linguistics, geography, etc.), regardless of whether it is explicitly familiar with it or not.

\subsection{Usage}
\label{sec:pythonlibrary}

To make the RDFGraphGen generator more easily available for the community, we packaged it as a Python library and published it online on PyPi\footref{fn:rdfgraphgen-pypi}. RDFGraphGen is available via the command: \verb|pip install rdf-graph-gen|, which downloads and installs it locally on the user's machine. Afterwards, RDFGraphGen can be used via the command-line, using the command: 

\begin{verbatim}
    rdfgen input-shapes.ttl output-graph.ttl scale-factor
\end{verbatim}

\noindent The command \verb|rdfgen| takes three arguments as input:
\begin{itemize}
    \item the file containing the input SHACL shapes graph, e.g., \texttt{input-shapes.ttl},
    \item the file that the generated synthetic RDF graph is to be written to, e.g., \texttt{output-graph.ttl}, and 
    \item the scale factor that determines the size of the generated RDF graph (\texttt{scale-factor}).
\end{itemize}

\noindent We optimized RDFGraphGen for generating large graphs --- data is generated in batches that are serialized in the output file when the batch size is reached, in order to continuously free up main memory. We also implemented concurrency to parallelize and speed up data generation for large graphs.

The full RDFGraphGen code, along with the CSV files containing pre-defined sets of values, examples of SHACL shape files and generated graph examples, are publicly available on GitHub\footref{fn:rdfgraphgen-github}. Using a CI/CD pipeline based on GitHub actions, we release a new version of the RDFGraphGen generator on PyPi when there are changes in the GitHub repository.

%%%%%%%%%%%%%%%%%%%%%%%%%%%%%%%%%%%%%%%%%%

\section{Example Use Case}
\label{sec:examples}

To showcase how RDFGraphGen works for specific SHACL shapes, this section presents an example of using the \texttt{schema:Person} class from schema.org. Other examples, from various domains using different ontologies are available on the project's GitHub page\footref{fn:rdfgraphgen-github}.

In this example, the input SHACL shapes graph (Listing 1) describes entities that are of type \texttt{schema:Person}. From the definition, each such entity is constrained to have a given name and a last name, or a full name; exactly one date of birth, which must be earlier then the potential death date; a gender with one of the available values; an email with a value constrained by a pattern; another using the \texttt{schema:email} property; a telephone number; a job title; an address, which has a specified complex type (it is an object, not a literal). Additionally, the address objects have a street address that is a string value; a postal code, which can be a string or an integer with values between 10,100--10,999.\\

% frame=lines,label=Input SHACL Shape
\begin{Verbatim}[fontsize=\small,frame=single,framesep=3mm,label=Listing 1. Input SHACL Shapes Graph,labelposition=topline,numbers=left]
ex:PersonShape a sh:NodeShape ;
    sh:targetClass schema:Person ;
    sh:xone ( [ sh:property [
                    sh:path schema:givenName ;
                    sh:datatype xsd:string ;
                    sh:minCount 1 ;
                ] ;
                sh:property [
                    sh:path schema:familyName ;
                    sh:datatype xsd:string ;
                    sh:minCount 1
                ]]
              [ sh:path schema:name ;
                sh:datatype xsd:string  ;
                sh:minCount 1 ] );
    sh:property [
        sh:path schema:birthDate ;
        sh:lessThan schema:deathDate ;
        sh:minCount 1 ;
        sh:maxCount 1 ] ;
    sh:property [
        sh:path schema:gender ;
        sh:in ( "female" "male" ) ;
        sh:maxCount 1 ] ;
    sh:property [ 
        sh:path ex:workMail ; 
        sh:pattern "^[a-z0-9]+\\.[a-z0-9]+@work\\.[a-z]{2,3}$" 
        ] ;
    sh:property [ 
        sh:path schema:email ; 
        sh:maxCount 1 ] ;
    sh:property [ sh:path schema:telephone ] ;
    sh:property [ sh:path schema:jobTitle ] ;
    sh:property [
        sh:path schema:address ;
        sh:node ex:AddressShape ;
        sh:minCount 1 ] .
    
ex:AddressShape a sh:NodeShape ;
    sh:class schema:PostalAddress ;
    sh:property [
        sh:path schema:streetAddress ;
        sh:datatype xsd:string ; ] ;
    sh:property [
        sh:path schema:postalCode ;
        sh:or ( [ sh:datatype xsd:integer ;
                  sh:minInclusive 10100 ;
                  sh:maxInclusive 10999  ]
                [ sh:datatype xsd:string ] ) ] .
\end{Verbatim}

After running the RDFGraphGen generator on the input SHACL shapes file and setting the parameter \texttt{scale-factor} to 2, we get the generated synthetic RDF graph (Listing 2).\\

%frame=lines,label=Output RDF Graph
\begin{Verbatim}[fontsize=\small,frame=single,framesep=3mm,label=Listing 2. Output RDF Graph,labelposition=topline,numbers=left]
<http://example.org/ns#Node100> a schema:Person ;
    ex:workMail "18b.dqjz8szqwne7nm@work.ekx" ;
    schema:address <http://example.org/ns#Node101> ;
    schema:birthDate "1965-07-07"^^xsd:date ;
    schema:deathDate "1989-07-07"^^xsd:date ;
    schema:email "barton_aldridge@gmail.com" ;
    schema:familyName "Aldridge" ;
    schema:gender "male" ;
    schema:jobTitle "bartender" ;
    schema:givenName "Barton" ;
    schema:telephone "647-466-552849" ;
    sh:description ex:PersonShape .

<http://example.org/ns#Node102> a schema:Person ;
    ex:workMail "sqq2.s7ojq@work.vab" ;
    schema:address <http://example.org/ns#Node103> ;
    schema:birthDate "1986-07-07"^^xsd:date ;
    schema:email "sarajanebenjamin@gmail.com" ;
    schema:gender "female" ;
    schema:jobTitle "psychologist" ;
    schema:name "Sarajane Benjamin" ;
    schema:telephone "722-279-0247032" ;
    sh:description ex:PersonShape .

<http://example.org/ns#Node101> a schema:PostalAddress ;
    schema:postalCode 10481 ;
    schema:streetAddress "no. 3 Lily st" ;
    sh:description ex:AddressShape .

<http://example.org/ns#Node103> a schema:PostalAddress ;
    schema:postalCode "6plO65qAPywG" ;
    schema:streetAddress "no. 1 Gillette ave" ;
    sh:description ex:AddressShape .
\end{Verbatim}    

When generating the synthetic RDF graph, given this input, RDFGraphGen produces an output such as the one presented in Listing 2. When it generates the name of each person entity, only one option from the \texttt{sh:xone} constraint can be selected (Line 3, Listing 1). For the first person (Line 1, Listing 2), the generator has created separate predicates for the given name (Line 10, Listing 2) and the family name (Line 7, Listing 2). For the second person (Line 14, Listing 2), the generator has generated a single predicate for the full name (Line 21, Listing 2), and the full name includes both the given name and the family name as a single value.

Furthermore, predicates related to dates have a date value for the object despite not explicitly containing a \texttt{sh:datatype} constraint in the description. The generator extracts information from the predicate names of \texttt{schema:birthDate} and \texttt{schema:deathDate} (Lines 17-18, Listing 1), determines that the objects' values in the generated RDF triples should be dates, and then generates them based on the constraints specified in the input SHACL shapes graph.

The email address values (\texttt{schema:email}) consist of the given name and the family name of the person (Lines 6 and 18, Listing 2). This has been specifically predefined in the generator when working with \texttt{schema:Person} entities and their email addresses. The other email address value (\texttt{ex:workMail}) has a definition that is not part of schema.org and therefore its value is constrained by the pattern in the SHACL graph (Line 27, Listing 1). The generator produces values that conform to the defined pattern (Lines 2 and 15, Listing 2).

The generated phone numbers follow a specific pattern. This is again based on the predicate name, \texttt{schema:telephone} (Line 32, Listing 1) and therefore the generator applies the predefined pattern constraint (Lines 11 and 22, Listing 2).

The address for each person is generated as a separate RDF entity (Lines 25 and 30, Listing 2) based on the SHACL shape in the input file, which constrains the address. Given that there are no explicit cardinality restrictions on the relation between people and their addresses, each person has exactly one address. So, given that \verb|scale-factor| was set to 2 when starting the generator, the output contains a total of 4 entities: 2 person entities (as top-level shapes) and 2 address entities (as shapes directly connected to a top-level shape).

% \bgroup
% \def\arraystretch{1}%  1 is the default, change whatever you need
\begin{table}[!ht]
\centering
\resizebox{\textwidth}{!}{\begin{tabular}{r||rr||rr||rr||rr}
% \toprule
% \toprule
\multicolumn{1}{c||}{} & \multicolumn{2}{c||}{\begin{tabular}[c]{@{}c@{}}Single \\ Simple Shape\end{tabular}} & \multicolumn{2}{c||}{\begin{tabular}[c]{@{}c@{}}Single \\ Complex Shape\end{tabular}} & \multicolumn{2}{c||}{\begin{tabular}[c]{@{}c@{}}Three \\ Simple Shapes\end{tabular}} & \multicolumn{2}{c}{\begin{tabular}[c]{@{}c@{}}Three \\ Complex Shapes\end{tabular}} \\ \hline
\multicolumn{1}{c||}{\begin{tabular}[c]{@{}c@{}}Scale\\Factor\end{tabular}} & \multicolumn{1}{c|}{\begin{tabular}[c]{@{}c@{}}RDF\\Triples\end{tabular}} & \multicolumn{1}{c||}{\begin{tabular}[c]{@{}c@{}}Time \\ (s)\end{tabular}} & \multicolumn{1}{c|}{\begin{tabular}[c]{@{}c@{}}RDF\\Triples\end{tabular}} & \multicolumn{1}{c||}{\begin{tabular}[c]{@{}c@{}}Time\\(s)\end{tabular}} & \multicolumn{1}{c|}{\begin{tabular}[c]{@{}c@{}}RDF\\Triples\end{tabular}} & \multicolumn{1}{c||}{\begin{tabular}[c]{@{}c@{}}Time\\(s)\end{tabular}} & \multicolumn{1}{c|}{\begin{tabular}[c]{@{}c@{}}RDF\\Triples\end{tabular}} & \multicolumn{1}{c}{\begin{tabular}[c]{@{}c@{}}Time\\(s)\end{tabular}} \\ 
\midrule
\midrule
1 & \multicolumn{1}{r|}{2} & 4 & \multicolumn{1}{r|}{9} & 4 & \multicolumn{1}{r|}{8} & 4 & \multicolumn{1}{r|}{13} & 4 \\ \hline
10 & \multicolumn{1}{r|}{20} & 4 & \multicolumn{1}{r|}{89} & 4 & \multicolumn{1}{r|}{80} & 4 & \multicolumn{1}{r|}{135} & 4 \\ \hline
100 & \multicolumn{1}{r|}{200} & 4 & \multicolumn{1}{r|}{862} & 4 & \multicolumn{1}{r|}{800} & 5 & \multicolumn{1}{r|}{1,401} & 5 \\ \hline
1,000 & \multicolumn{1}{r|}{2,000} & 5 & \multicolumn{1}{r|}{8,477} & 5 & \multicolumn{1}{r|}{8,000} & 5 & \multicolumn{1}{r|}{13,939 } & 5 \\ \hline
10,000 & \multicolumn{1}{r|}{20,000} & 8 & \multicolumn{1}{r|}{84,975} & 8 & \multicolumn{1}{r|}{80,000} & 8 & \multicolumn{1}{r|}{139,969} & 9 \\ \hline
100,000 & \multicolumn{1}{r|}{200,000} & 10 & \multicolumn{1}{r|}{850,313} & 24 & \multicolumn{1}{r|}{800,000} & 16 & \multicolumn{1}{r|}{1,400,109} & 30 \\ \hline
1,000,000 & \multicolumn{1}{r|}{2,000,000} & 35 & \multicolumn{1}{r|}{8,504,410} & 164 & \multicolumn{1}{r|}{8,000,000} & 102 & \multicolumn{1}{r|}{14,004,068} & 244 \\ \hline
10,000,000 & \multicolumn{1}{r|}{20,000,000} & 287 & \multicolumn{1}{r|}{85,042,729} & 1,608 & \multicolumn{1}{r|}{80,000,000} & 1,352 & \multicolumn{1}{r|}{140,044,876} & 2,335 \\ \hline
100,000,000 & \multicolumn{1}{r|}{200,000,000} & 2,871 & \multicolumn{1}{r|}{850,497,903} & 17,312 & \multicolumn{1}{r|}{800,000,000} & 9,888 & \multicolumn{1}{r|}{1,371,272,422} & 22,734 \\ % \hline
% 1,000,000,000 & \multicolumn{1}{r|}{TBD} & TBD & \multicolumn{1}{r|}{TBD} & TBD & \multicolumn{1}{r|}{TBD} & TBD & \multicolumn{1}{r|}{TBD} & TBD \\
\bottomrule
\bottomrule
\end{tabular}}
\vspace{5 pt}
\caption{Performance Evaluation of RDFGraphGen}
\label{tab:performance}
\end{table}
% \egroup

\section{Performance Evaluation}
\label{sec:performance-evaluation}

To evaluate the performance of RDFGraphGen, we measured the processing time for generating some specific RDF datasets. The experiment suite is available on the project GitHub repository\footref{fn:rdfgraphgen-github}.

We tested three dimensions: scale factor (from 1 to 100,000,000), number of shapes in the input SHACL graph (1 and 3), and SHACL shape complexity per input shape (from a single string predicate, up to 4 properties in varying datatypes, lengths, cardinalities, etc.). All experiments were run on a MacBook Pro with an Apple M4 chip, 10-core CPU, 24GB main memory and 120GB/s memory bandwidth.

\begin{figure}[!ht]
    \centering
    \includegraphics[width=1\linewidth]{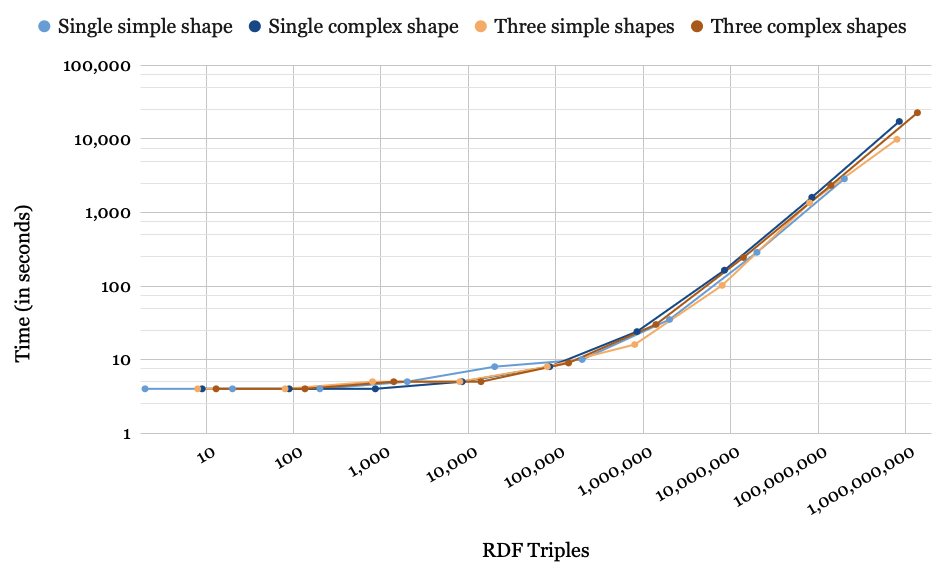}
    \caption{Generation Time based on the Number of Generated RDF Triples}
    \label{fig:generation-time-chart}
\end{figure}

Our experiments, presented in Table \ref{tab:performance}, show that RDFGraphGen generates synthetic RDF graphs in a reasonably good timeframe. %, comparable to other similar tools (e.g. GAIA \cite{raynaud2016generic}). 
The generation time slightly depends on the complexity of the input SHACL shapes but is more strongly correlated with the number of RDF triples being generated (Figure \ref{fig:generation-time-chart}). The concurrent implementation of RDFGraphGen allows it to use all available CPUs on a given machine and scale efficiently. In our tests RDFGraphGen automatically ran with 10 processes in parallel.%, due to the 10 available CPUs.

\section{Conclusion and Future Work}
\label{sec:conclusion}

In this paper, we introduced RDFGraphGen --- a general-purpose, domain-agnostic synthetic RDF graph generator based on SHACL shapes. To the best of our knowledge, this is the first RDF generator based on SHACL constraints. The synthetic RDF knowledge graphs generated by RDFGraphGen can be used in many different scenarios in the software development cycle: application testing, algorithm testing, application benchmarking, software quality control, training of machine learning models, etc. 

RDFGraphGen is domain-independent: the generated RDF graph contains data from the domain that is described in the source SHACL shapes file. Additionally, RDFGraphGen has the ability to use pre-collected values for schema.org classes and predicates and can be extend to include values for any ontology. This allows for the generator to be fine-tuned and adapted to the needs of specific use cases and domains.

In the future, we plan to extend the support for human-friendly literal values for an increased number of ontologies, classes, and properties in a manner similar to what we did with a subset of schema.org. To make our finding reproducible, and enable others to easily use RDFGraphGen, we have published it as open-source, under the MIT license.

%%%%%%%%%%%%%%%%%%%%%%%%%%%%%%%%%%%%%%%%%%

\section*{Acknowledgment}
\label{sec:ack}

The work in this paper was partially financed by the Faculty of Computer Science and Engineering, Ss. Cyril and Methodius University in Skopje.

%%%%%%%%%%%%%%%%%%%%%%%%%%%%%%%%%%%%%%%%%%

\bibliographystyle{plainnat}
\bibliography{bibliography}

\end{document}